\documentclass[aps,prd,superscriptaddress,twoside,twocolumn,showpacs]{revtex4}
\usepackage{amsmath,amssymb}
\usepackage{graphicx}

\begin{document}
\title{New determination of the mass of the $\boldsymbol{\eta}$ meson at COSY-ANKE}
\author{P.~Goslawski}
\email{paul.goslawski@uni-muenster.de}
\affiliation{Institut f\"ur Kernphysik, Universit\"at M\"unster, D-48149 M\"unster, Germany}
\author{A.~Khoukaz}
\email{khoukaz@uni-muenster.de}
\affiliation{Institut f\"ur Kernphysik, Universit\"at M\"unster, D-48149 M\"unster, Germany}
\author{S.~Barsov}
\affiliation{High Energy Physics Department, Petersburg Nuclear Physics Institute, RU-188350 Gatchina, Russia}
\author{I.~Burmeister}
\affiliation{Institut f\"ur Kernphysik, Universit\"at M\"unster, D-48149 M\"unster, Germany}
\author{D.~Chiladze}
\affiliation{Institut f\"ur Kernphysik and J\"ulich Centre for Hadron Physics, Forschungszentrum J\"ulich, D-52425 J\"ulich, Germany}
\affiliation{High Energy Physics Institute, Tbilisi State University, GE-0186
Tbilisi, Georgia}
\author{S.~Dymov}
\affiliation{Physikalisches Institut, Universit{\"a}t Erlangen-N\"urnberg, D-91058 Erlangen, Germany}
\affiliation{Laboratory of Nuclear Problems, JINR, RU-141980 Dubna, Russia}
\author{R.~Gebel}
\affiliation{Institut f\"ur Kernphysik and J\"ulich Centre for Hadron Physics, Forschungszentrum J\"ulich, D-52425 J\"ulich, Germany}
\author{M.~Hartmann}
\affiliation{Institut f\"ur Kernphysik and J\"ulich Centre for Hadron Physics, Forschungszentrum J\"ulich, D-52425 J\"ulich, Germany}
\author{A.~Kacharava}
\affiliation{Institut f\"ur Kernphysik and J\"ulich Centre for Hadron Physics, Forschungszentrum J\"ulich, D-52425 J\"ulich, Germany}
\author{P.~Kulessa}
\affiliation{H.~Niewodniczanski Institute of Nuclear Physics PAN, PL-31342 Cracow, Poland}
\author{A.~Lehrach}
\affiliation{Institut f\"ur Kernphysik and J\"ulich Centre for Hadron Physics, Forschungszentrum J\"ulich, D-52425 J\"ulich, Germany}
\author{B.~Lorentz}
\affiliation{Institut f\"ur Kernphysik and J\"ulich Centre for Hadron Physics, Forschungszentrum J\"ulich, D-52425 J\"ulich, Germany}
\author{R.~Maier}
\affiliation{Institut f\"ur Kernphysik and J\"ulich Centre for Hadron Physics, Forschungszentrum J\"ulich, D-52425 J\"ulich, Germany}
\author{T.~Mersmann}
\affiliation{Institut f\"ur Kernphysik, Universit\"at M\"unster, D-48149 M\"unster, Germany}
\author{M.~Mielke}
\affiliation{Institut f\"ur Kernphysik, Universit\"at M\"unster, D-48149 M\"unster, Germany}
\author{S.~Mikirtychiants}
\affiliation{Institut f\"ur Kernphysik and J\"ulich Centre for Hadron Physics, Forschungszentrum J\"ulich, D-52425 J\"ulich, Germany}
\affiliation{High Energy Physics Department, Petersburg Nuclear Physics Institute, RU-188350 Gatchina, Russia}
\author{H.~Ohm}
\affiliation{Institut f\"ur Kernphysik and J\"ulich Centre for Hadron Physics, Forschungszentrum J\"ulich, D-52425 J\"ulich, Germany}
\author{M.~Papenbrock}
\affiliation{Institut f\"ur Kernphysik, Universit\"at M\"unster, D-48149 M\"unster, Germany}
\author{D.~Prasuhn}
\affiliation{Institut f\"ur Kernphysik and J\"ulich Centre for Hadron Physics, Forschungszentrum J\"ulich, D-52425 J\"ulich, Germany}
\author{T.~Rausmann}
\affiliation{Institut f\"ur Kernphysik, Universit\"at M\"unster, D-48149 M\"unster, Germany}
\author{V.~Serdyuk}
\affiliation{Laboratory of Nuclear Problems, JINR, RU-141980 Dubna, Russia}
\author{H.~Stockhorst}
\affiliation{Institut f\"ur Kernphysik and J\"ulich Centre for Hadron Physics, Forschungszentrum J\"ulich, D-52425 J\"ulich, Germany}
\author{H.~Str\"oher}
\affiliation{Institut f\"ur Kernphysik and J\"ulich Centre for Hadron Physics, Forschungszentrum J\"ulich, D-52425 J\"ulich, Germany}
\author{A.~T\"aschner}
\affiliation{Institut f\"ur Kernphysik, Universit\"at M\"unster, D-48149 M\"unster, Germany}
\author{Yu.~Valdau}
\affiliation{Institut f\"ur Kernphysik and J\"ulich Centre for Hadron
Physics, Forschungszentrum J\"ulich, D-52425 J\"ulich,
Germany}\affiliation{Helmholtz-Institut f\"ur Strahlen- und Kernphysik,
Universit\"at Bonn, D-53115 Bonn, Germany}
\author{C.~Wilkin}
\affiliation{Physics and Astronomy Department, UCL, Gower Street, London WC1E 6BT, U.K.}

\date{\today}

\begin{abstract}
A value for the mass of the $\eta$ meson has been determined at
the COSY-ANKE facility through the measurement of a set of
deuteron laboratory beam momenta and associated $^3$He
center-of-mass momenta in the $dp\to{}^{3}$He$\,X$ reaction.
The $\eta$ was then identified by the missing-mass peak and the
production threshold determined. The individual beam momenta
were fixed with a relative precision of $3 \times 10^{-5}$ for
values around 3~GeV/$c$ by using a polarized deuteron beam and
inducing an artificial depolarizing spin resonance, which
occurs at a well-defined frequency. The final-state momenta in
the two-body $dp\to{}^{3}$He$\,\eta$ reaction were investigated
in detail by studying the size of the $^3$He momentum ellipse
with the forward detection system of the ANKE spectrometer.
Final alignment of the spectrometer for this high precision
experiment was achieved through a comprehensive study of the
$^3$He final-state momenta as a function of the center-of-mass
angles, taking advantage of the full geometrical acceptance.
The value obtained for the mass,
$m_{\eta}=(547.873\pm0.005_{\text{stat}}\pm0.027_{\text{syst}})$~MeV/$c^2$,
is consistent and competitive with other recent measurements,
in which the meson was detected through its decay products.
\end{abstract}


\maketitle

\section{Introduction}
\label{sec:Intro} The precise value of the mass of the $\eta$
meson has been the subject of intense debate for several years.
The situation seems to have been finally resolved with the
publication of four experiments that obtained consistent
results to high
accuracy~\cite{LAI2002,AMB2007,MAMI2012,MIL2007}. In all these
new experiments the meson was cleanly identified through one or
more of its decay modes, $\pi^0\pi^0\pi^0$~\cite{LAI2002},
$\gamma\gamma$~\cite{AMB2007}, or both~\cite{MAMI2012}, or
these plus $\pi^+\pi^-\pi^0$ and
$\pi^+\pi^-\gamma$~\cite{MIL2007}. Taking only decay
experiments into account, the Particle Data Group now quote
their ``best'' estimate of the mass as being
 $m_{\eta} = (547.853 \pm 0.024)$~MeV/$c^2$~\cite{PDG2010}.
Experiments in which the $\eta$ meson was identified through a
missing-mass peak in a hadronic production reaction have all
reported a lower value for the mass, typically by about
$0.5$~MeV/$c^2$. This was the case for the $\pi^-p\to n\eta$
reaction, where the beam momentum was determined
macroscopically to high precision using the floating wire
technique~\cite{DUA1974}. In the two experiments where the $dp
\, (pd)\to{}^{3}$He$\,\eta$ reaction was
used~\cite{PLO1992,ABD2005}, the beam momentum was measured by
studying other two-body reactions with known final masses. One
possible concern might be whether the background under the
$\eta$ peak could have been slightly distorted by a strong
coupling of, for example, $\eta\,^3$He$\,\rightleftarrows
\pi\pi\,^3$He. Alternatively, the beam momenta could have been
poorly determined, though this was done using different
techniques for the three
experiments~\cite{DUA1974,PLO1992,ABD2005}, or the
spectrometers aligned insufficiently well.

The situation can only be clarified through the performance of
a much more precise missing-mass experiment~\cite{KHO2007}. It is the purpose
of the present paper to provide results of such an
investigation of the two-body $dp\to{}^{3}$He$\,\eta$ reaction
near threshold that is comparable in accuracy with those that
studied the $\eta$ decay
products~\cite{LAI2002,AMB2007,MAMI2012,MIL2007}. For this to
be successfully achieved, it is necessary (i) to
establish the beam momentum, (ii) to identify well the
$\eta$ meson from the $dp\to{}^{3}$He$\,X$ missing-mass peak,
and (iii) to establish the increase of the final-state
momentum with excess energy
$Q=\sqrt{s}-(m_{^3\text{He}}-m_{\eta})c^2$ above the
$^{3}$He$\,\eta$ reaction threshold with high accuracy. Here
$\sqrt{s}$ is the total center-of-mass (c.m.) energy. A data set
of beam momenta and associated final-state momenta near
threshold then permits the determination of the production
threshold and hence the $\eta$ meson mass.

In order to obtain a clean identification of the $\eta$ meson
from a missing-mass peak in a $dp\to{}^{3}$He$\,X$ reaction, it
is important that $\eta$ production be very strong in the
near-threshold region. Here we are particularly fortunate in
that the $dp\to{}^{3}$He$\,\eta$ total cross section rises
within the first 1~MeV above threshold to its plateau value of
$\approx 400$~nb and then remains nearly constant up to
$Q=100$~MeV~\cite{MER2007,RAU2009}. This allows one to collect
similar statistics over a wide excess energy range, even very
close to threshold, without expending excessive measuring time.
Furthermore, it has been demonstrated that the ANKE facility is
well suited for measuring the total and differential cross
sections of the $dp\to{}^{3}$He$\,\eta$ reaction. The
spectrometer has full geometrical acceptance for excess
energies $Q < 15$~MeV. The multipion background under the
$\eta$ peak varies smoothly with $Q$, making a robust
subtraction of this background possible~\cite{MER2007,RAU2009}.

The extrapolation of data to identify the production threshold
requires a precise measurement of the increase of the $^3$He c.m.
momentum as a function of the beam momentum. Due to resolution
or smearing effects, which are always present in any real
detector, the reconstructed momenta can be shifted compared to
the true ones. To understand and compensate for such effects, a
careful study of the entire ANKE spectrometer is needed. After
calibrating the spectrometer by measuring a variety of other
nuclear reactions, the requisite precision was achieved by
demanding that the magnitude of the true $^3$He momentum from
the $dp\to{}^{3}$He$\,\eta$ reaction should be identical in all
directions in the c.m. frame. This check was only possible
because of the 100\% angular acceptance of ANKE for the
reaction of interest.

We have previously described in full the measurement of the
beam momenta via the induced spin depolarization
technique~\cite{GOS2010}. The background subtraction that
allows the extraction of the $\eta$ meson signal from the
$dp\to{}^{3}$He$\,X$ reaction is essentially identical to that
of our earlier work~\cite{MER2007}. Therefore, most of the
emphasis here will be on describing the fine calibrations of
the spectrometer required to get the necessary precision in the
final $^3$He momenta. After outlining the method for
determining the $\eta$ mass in Sec.~\ref{sec:Method}, the beam
momentum determination is briefly summarized in
Sec.~\ref{sec:BeamMom}. The standard calibration of the ANKE
forward detector system is given in Sec.~\ref{sec:ANKE}.
Section~\ref{sec:FM} discusses how to exploit the full
geometrical acceptance of a two-body reaction, such as
$dp\to{}^{3}$He$\,\eta$, to verify and improve the alignment of
the spectrometer. Results and estimated uncertainties are
presented and it is shown how resolution effects can influence
the measurement of the final-state momenta and hence the
missing mass.

The results of the determinations of the beam and final-state
momenta are brought together in Sec.~\ref{sec:EtaMass} so that
a reliable extrapolation to the $\eta$ threshold can be made.
In this way, a value for the mass of the $\eta$ meson  was
obtained with small systematic and negligible statistical
errors. The comparison with the results of other experiments is
made in our conclusions of Sec.~\ref{sec:Conc}. The good
agreement with other precision measurements was only achieved
through the exploitation of the large angular acceptance of
ANKE for this $\eta$-production reaction, a feature that was
not available in earlier missing-mass experiments.

\section{Method for the determination of the $\boldsymbol{\eta}$ meson mass}
\label{sec:Method}

By studying the two-body $dp\to{}^{3}$He$\,\eta$ reaction in a
fixed-target experiment, the $\eta$ mass can be determined by
measuring only the momenta of the deuteron beam and the
recoiling $^3$He. In a standard missing-mass experiment, the
$\eta$ mass is extracted by measuring the relevant kinematic
variables at a single fixed energy. This requires both a
precise calibration of the detector, for a correct
determination of the $^3$He momentum, and an accurate
measurement of the deuteron beam momentum.

A much more effective way to measure the $\eta$ mass relies on
the determination of the production threshold by investigating
the change of the final-state momentum as a function of the
beam momentum. This requires the kinematics to be measured at
several different energies close to threshold. For the
$dp\to{}^{3}$He$\,\eta$ reaction, the final-state momentum
$p_f$ in the c.m. frame
\begin{equation}
	\label{eq:pf}
	p_f(s) = \frac{\sqrt{ \left[s - \left(m_{^{3}\text{He}} + m_{\eta} \right)^2 \right]
	\left[s - \left(m_{^{3}\text{He}} - m_{\eta} \right)^2 \right]    }} {2\sqrt{s}} \, ,
\end{equation}
is a very sensitive function of the $\eta$ mass and the total
energy $\sqrt{s}$. The latter is completely fixed by the masses
of the initial particles and the laboratory momentum, $p_d$, of
the deuteron beam;
\begin{equation}
	\label{eq:s}
	s = 2m_p \sqrt{m_d^2 + p_d^2} + m_d^2 +m_p^2.
\end{equation}

The final-state momentum depends only on the beam momentum, the
$\eta$ mass and other well-measured masses~\cite{NIST}. If one
can fix the production threshold, $p_f(s)=0$, the $\eta$ mass
can then be determined from knowledge of $p_d$. The precision
is enhanced because in this region $dm_{\eta}/dp_d\approx
0.24/c$.

\begin{figure}[hbt]
\centering
\includegraphics[width=1.0\linewidth]{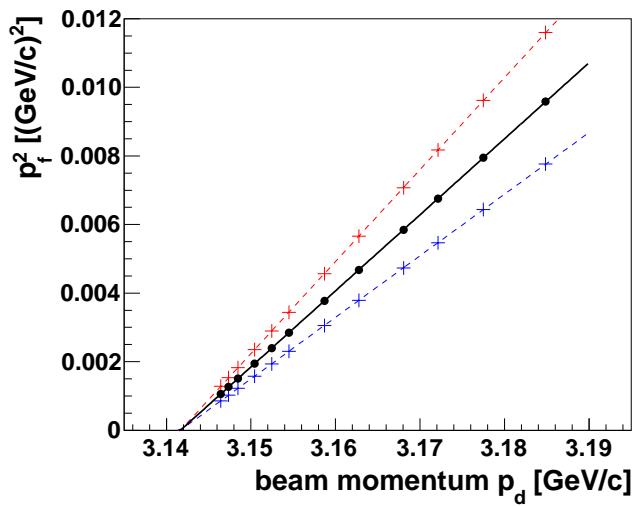}
\caption{\label{fig1} (color online) Identification of the
$dp\rightarrow\,^{3}\text{He}\,\eta$ production threshold by
studying the relationship between the square of the final-state
momentum $p_f$ and the deuteron beam momentum $p_d$. The twelve
Monte Carlo points are compared with the shape expected from
the kinematics (solid line). Although this shape depends
slightly on $m_{\eta}$, it is essentially a straight line over
the range shown. The stability of the threshold extrapolation
is illustrated by the dashed red and blue lines, where the
measurements of $p_f$ are scaled by arbitrary constants $S=0.9$
and $1.1$, respectively.}
\end{figure}

An obvious advantage of the threshold determination method is
that it does not require a perfect spectrometer calibration.
This is illustrated in Fig.~\ref{fig1} using Monte Carlo
simulated data at twelve different excess energies in the range
of $Q = 1 - 11$~MeV and assuming some fixed value of the $\eta$
mass. Although the analysis was performed using
Eq.~\eqref{eq:pf} (solid line), to first order $p_f^2$ depends
linearly on $p_d$ near threshold.

To reduce the sensitivity of the method to systematic errors in
the spectrometer calibration, an additional scaling factor $S$
is introduced which multiplies the right-hand side of
Eq.~\eqref{eq:pf}. Such a factor will occur in a real
experiment, e.g., through minor inaccuracies in the
determination of the interaction vertex relative to the
detection system. If one then used data at a single excess
energy, this would lead to an error in the determination of the
missing mass. However, by measuring over a range of excess
energies, such a scaling factor would not affect the value of
$p_d$ corresponding to the production threshold that is used in
the mass determination. This effect is illustrated by the
dashed lines in Fig.~\ref{fig1}, where 10\% changes in the
scaling factor were considered.

\section{Beam momentum determination}
\label{sec:BeamMom}

A method to determine the beam momentum to high precision in a
storage ring was developed at the electron-positron machine
VEPP-2M at Novosibirsk~\cite{DER1980}. The technique uses the
spin dynamics of a polarized beam and relies on the fact that
depolarizing resonances occur at well-defined frequencies,
which, apart from the gyromagnetic anomaly, depend only upon a
particle's speed. By depolarizing a polarized beam with an
artificial spin resonance induced by a horizontal radio
frequency magnetic field of a solenoid, the value of the beam
momentum can be measured to high accuracy. The frequency of
such an artificial resonance $f_r$ is fixed in terms of the
machine frequency $f_0$ by the relativistic $\gamma$ factor of
the particle. Since frequencies can be routinely measured with
a relative precision $\approx 10^{-5}$, an accurate value of
$\gamma$ and hence momentum can be deduced. When this method
was used for the first time at COSY with a vector-polarized
deuteron beam, the precision achieved was more than an order of
magnitude better than that of conventional
methods~\cite{GOS2010}.

In the COSY-ANKE experiment, twelve closely-spaced beam momenta
above the $dp\rightarrow\,^{3}\text{He}\,\eta$ threshold were
divided alternately into two so-called supercycles that
involved up to seven different machine settings. In addition to
six momenta above threshold, one was included below threshold
in both supercycles to provide the background description. Each
supercycle covered an excess energy range from $\approx 1$~MeV
to $\approx 10$~MeV. The beam momenta were measured both before
and after five days of data-taking in order to study systematic
effects.

In both supercycles there were collective shifts of the spin
resonance frequencies of $\approx 12-17$~Hz, which probably
originated from changes in the orbit length in COSY of about
3~mm. The set of beam momenta from the first supercycle were
found to decrease slightly over the data-taking time whereas
those of the second supercycle showed the inverse behavior.
Average values of the measurements, before and after
data-taking, were calculated for each of the twelve beam
momenta and these were used in the threshold fits. The maximum
uncertainty for these mean values was conservatively evaluated
as $\pm 164$~keV/$c$. It is reasonable to assume that the
orbits changed linearly in time to give a uniform probability
distribution of true momenta over this interval. The systematic
uncertainty of the averaged beam momenta is therefore estimated
to be $\Delta p_{d, \rm{syst}}  = 95$~keV/$c$ (rms). Evidence
in favor of the approach adopted here is found by comparing the
results obtained with the two supercycles, which are discussed
in Sec.~\ref{sec:EtaMass}.

The statistical uncertainty in one of the twelve beam momenta
in the range 3.1\,--\,3.2~GeV/$c$ is dominated by that of the
revolution frequency $f_0$. The measured revolution frequency
deviated randomly by up to $\pm 6$~Hz from measurement to
measurement. The rms uncertainty in this frequency, 3.5~Hz,
corresponds to a statistical uncertainty in the beam momentum
of $\Delta p_{d, \rm{stat}} = 29$~keV/$c$.

In total, the twelve beam momenta were measured with an overall
accuracy of $\Delta p_d/p_d = 3 \times 10^{-5}$. This is
sufficient to satisfy the needs of a competitive $\eta$ mass
measurement.

\section{The ANKE setup}
\label{sec:ANKE}

\subsection{The ANKE magnetic spectrometer}
\label{subsec41}

\begin{figure*}[htb]				
\centering
\includegraphics[width=0.9\linewidth]{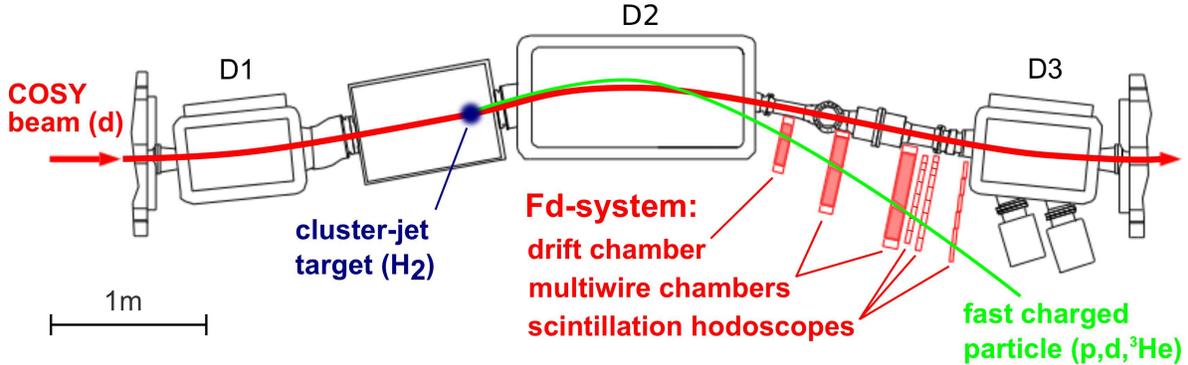}
\caption{\label{fig2} (color online)  The ANKE setup used for
the determination of the $\eta$ mass. The COSY deuteron beam
hits the hydrogen cluster-jet target and the charged particles
produced at small angles are separated by the D2 spectrometer
magnet to be detected by the ANKE Fd-system.}
\end{figure*}

The experiment was performed at the Cooler-Synchrotron of the
Forschungszentrum J\"ulich~\cite{MAI1997} using the ANKE
magnetic spectrometer~\cite{BAR2001} that is located at an
internal target station of the storage ring. ANKE includes
three dipole magnets (D1, D2, D3) that deflect the COSY beam
through a chicane in the ring (see Fig.~\ref{fig2}). This
allows particles produced at small angles, which would normally
escape down the beam pipe, to be detected and this is of
particular importance for near-threshold reactions. D1 deflects
the circulating beam by an angle off its straight path onto a
cluster-jet target~\cite{KHO1999}, the spectrometer dipole
magnet D2 separates the produced particles from the beam for
momentum analysis, and D3 leads the unscattered particles back
onto the nominal orbit. Although ANKE is equipped with a
variety of detectors, only the Forward system (Fd) was used for
the $\eta$ mass measurement. This consists of one drift
chamber, two multiwire proportional chambers, and three layers
of scintillation hodoscopes~\cite{CHI2002,DYM2004}. The
location of the various elements of the experimental setup is
illustrated in Fig.~\ref{fig2}.

The design requires the D2 magnet and the forward system to be
placed jointly on a moveable platform. The deflection angle of
the beam, $\alpha$, depends on the position of the platform.
The angle ($0^{\circ}\leq \alpha \leq 10.6^{\circ}$), the
magnetic field strength of D2 ($\leq 1.57$~T), and the beam
momentum cannot be chosen independently of each other. By
adjusting these three parameters it is possible to increase the
geometrical acceptance. The deflection angle, which may differ
slightly from the nominal one, was determined to be
$5.8^{\circ}$.

The tracks of fast charged particles detected in the ANKE
Fd-system can be traced back through the precisely known
magnetic field to the interaction point and this leads to a
momentum reconstruction for registered particles.

\subsection{The standard Fd-system calibration}
\label{subsec42}

In addition to the three parameters mentioned, i.e., the
magnetic field, deflection angle, and interaction point, the
drift and wire chamber positions also have to be determined
with high accuracy, and this requires a precise calibration of
the detection system. In the standard procedure~\cite{DYM2009},
the positions of the drift and wire chambers on the movable
platform are first aligned by using data taken at the beginning
of the beam time with a deflection angle of  $0^{\circ}$ and no
magnetic field in D2. The ejectiles then move on straight
tracks starting from the nominal interaction point in the
overlap region of the COSY beam and the cluster-jet target.
From an analysis of these tracks, the position of the
Fd-detector relative to the D2 magnet and the target can be
reconstructed and compared to direct measurements.

After making this first alignment, the global positions of the
drift and wire chambers are defined by that of the moveable
Fd-system platform. Although the positions of the platform and
interaction vertex are already known by direct measurement, a
much more precise determination of their values is possible by
investigating a series of reference reactions at all the
energies used in the $\eta$-mass determination. These are:
\begin{enumerate}
	\item Small angle $dp \rightarrow dp$ elastic
scattering, with the fast forward deuteron being detected,
	\item Large angle $dp \rightarrow dp$ elastic
scattering, with both final-state particles being detected,
	\item $dp \rightarrow ppn$ charge-exchange, with two
fast protons being detected, and
	\item $dp \rightarrow \, ^{3}\text{He}\,\pi^{0}$, with
the $^3$He nucleus being detected.
\end{enumerate}

Deuteron-proton elastic scattering in the backward hemisphere
allows one to verify energy-momentum conservation in the
reconstructed four-momenta. For the other reactions, the
minimization of the deviation of the missing mass from the
expected value was used to fix the positions of the interaction
vertex and the Fd-system platform. With the parameters
extracted at thirteen different beam energies, the missing
masses were reconstructed to an accuracy of $\approx
3$~MeV/$c^2$ for all four channels studied. Though good, it is
manifestly insufficient for a competitive determination of the
$\eta$ mass and the more refined technique discussed in the
following section is needed.

\section{The $\boldsymbol{^3\textrm{He} \, \eta}$ final-state momentum analysis}
\label{sec:FM}

\subsection{Fine calibration using the kinematics of the
$\boldsymbol{dp \rightarrow \, ^{3}\text{He}\,\eta}$ reaction}
\label{subsec:FM1}

The calibration method described in Sec.~\ref{subsec42}, i.e.,
the study of kinematic variables in measured reactions with
known masses, is standard for magnetic spectrometers and was
used in such a way in other $\eta$ missing-mass
experiments~\cite{PLO1992,ABD2005}. Unlike these experiments,
the ANKE facility has full geometrical acceptance for the
$^3$He from the $dp \rightarrow \, ^{3}\text{He}\,\eta$
reaction up to an excess energy $Q \approx 15$~MeV. By taking
advantage of this feature and studying the dependence of the
final-state momentum on the $^3$He c.m. angles, the standard
calibration could be significantly improved.

\begin{figure}[hbt]
\centering
\includegraphics[width=1.0\linewidth]{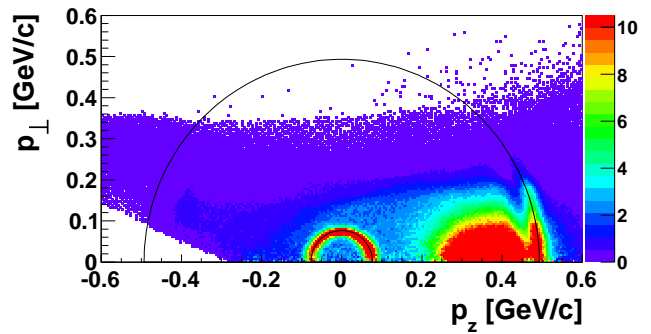}
\caption{\label{fig3} (color online) The magnitude of the
reconstructed transverse c.m. momentum $p_{\bot}$ in the
$dp\rightarrow\,^{3}\text{He}\,X$ reaction plotted against the
longitudinal c.m. component $p_z$ at an excess energy $Q=6.3$~MeV
with respect to the $\eta$ threshold. For better visualization
of the angular distribution, each event is weighted with a
factor $1/p_{\bot}$. The small and large circles correspond to
the kinematic loci for the $dp\rightarrow\,^{3}\text{He}\,\eta$
and $dp\rightarrow\,^{3}\text{He}\,\pi^{0}$ reactions,
respectively. ANKE covers the full solid angle for $\eta$
production near threshold whereas, for pions, only the forward
$^3$He are detected.}
\end{figure}

For a two-body reaction at a fixed center-of-mass energy, the
final-state momenta $p_f$ are distributed on a sphere in the c.m.
frame with a constant radius given by Eq.~\eqref{eq:pf}. We
take the $z$-direction to lie along that of the beam, $y$ to be
defined by the upward normal to the COSY ring, and
$\hat{x}=\hat{y}\times\hat{z}$. The relation of
Eq.~\eqref{eq:pf} can be visualized in a simplified way by
plotting the magnitude of the transverse momentum, $p_{\bot} =
\sqrt{p_x^2 + p_y^2}$, versus the longitudinal momentum $p_z$
for reconstructed events, as shown in Fig.~\ref{fig3}. For a
better visualization of the angular distribution, each event is
weighted with a factor $1/p_{\bot}$. The expected kinematic
loci for $dp \rightarrow \, ^{3}\text{He}\,\eta$ and $dp
\rightarrow \, ^{3}\text{He}\,\pi^{0}$ are shown by solid
lines. It is clear that the numbers of events vary along the
circles but the method used here only depends upon the position
of a kinematic curve and not on its population. In addition to
single meson production, there is a large accumulation of
events near the forward direction for $p_f\approx 350$~MeV/$c$
and these correspond to two-pion production in the $dp
\rightarrow \, ^{3}\text{He}(\pi\pi)^{0}$ reaction.

The principle of the refined spectrometer calibration is the
requirement that the momentum sphere should be completely
symmetric in $p_x$, $p_y$, and $p_z$. It is therefore necessary
to study the reconstructed momentum $p_f$ carefully as a
function of the polar and azimuthal angles $\vartheta$ and
$\phi$. This requires a clean separation of the $^3$He$\, \eta$
signal from the background and this is the first step in the
analysis.

\begin{figure}[htb]
\centering
\includegraphics[width=1.0\linewidth]{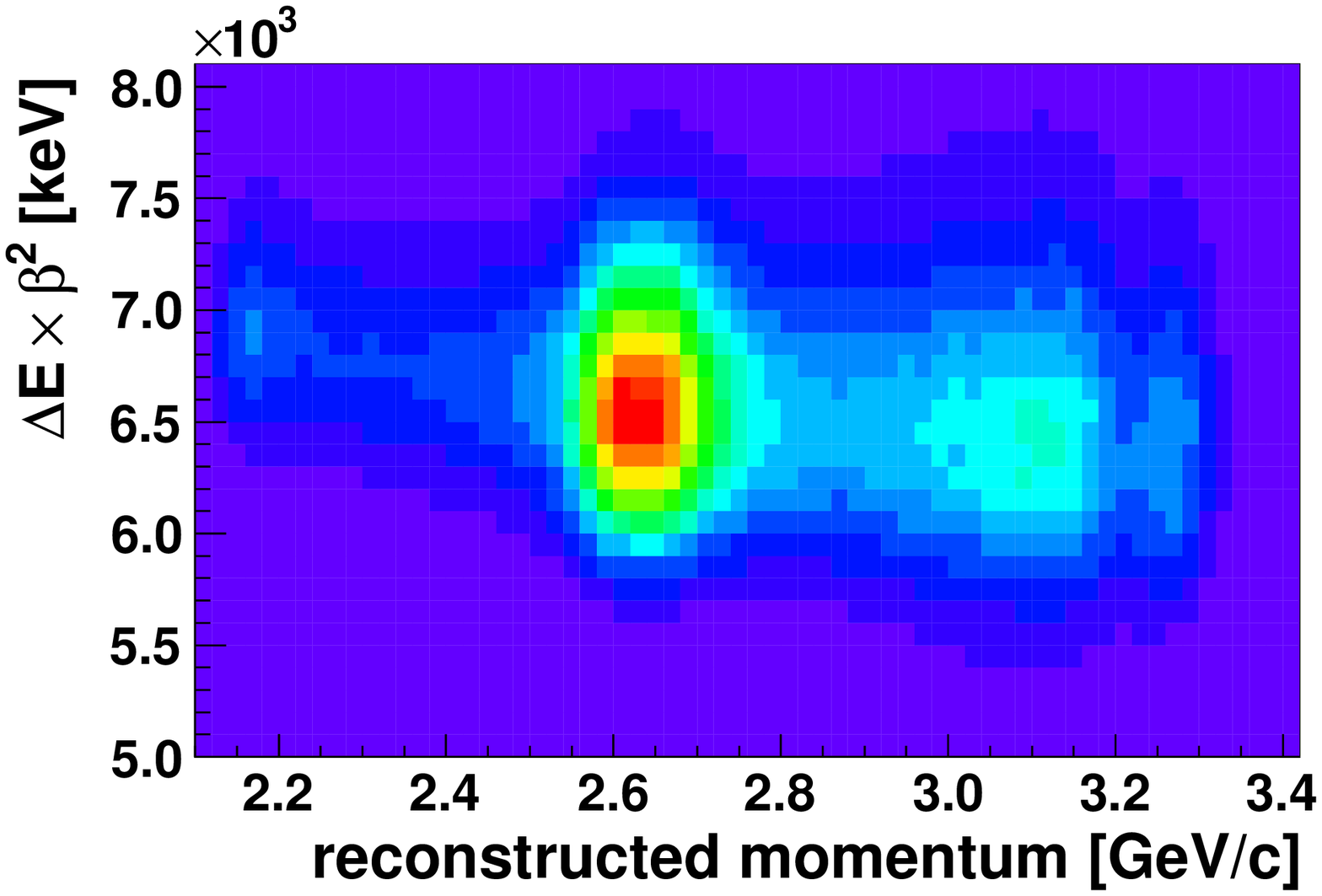}
\includegraphics[width=1.0\linewidth]{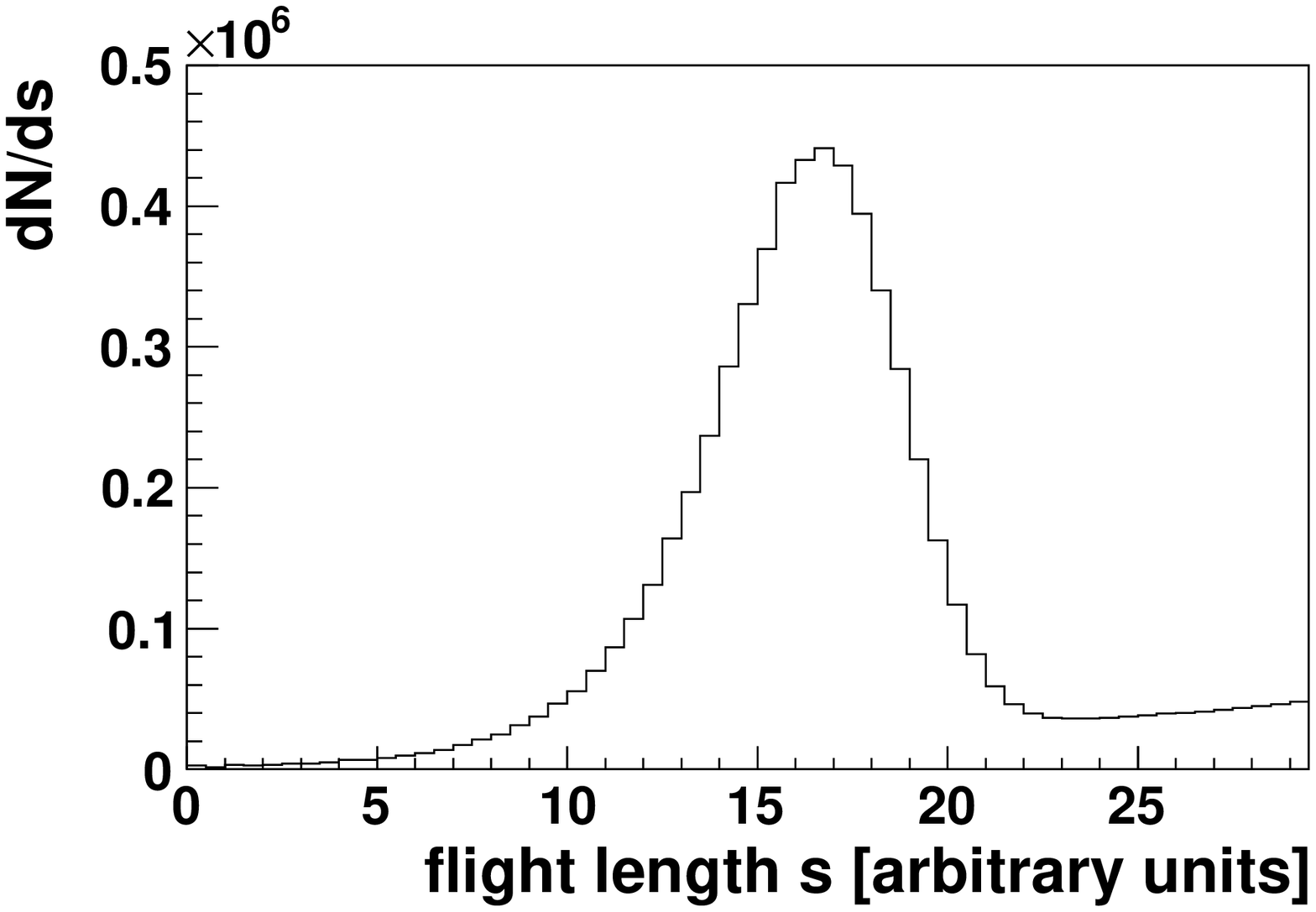}
\caption{\label{fig4} (color online) (a) Two-dimensional
distribution of energy loss times the square of the particle
velocity versus its laboratory momentum. A projection of this
onto the $y$-axis gives a peak with little background. Data
within a $\pm 6\sigma$ cut were retained for further analysis.
(b) A typical flight length distribution for events selected
from panel (a). Data within a $\pm 3\sigma$ range were used in
the subsequent analysis.}
\end{figure}

In the event selection, $^3$He particles were identified and
the raw background, consisting mainly of deuterons and protons
from $dp$ elastic scattering and protons from deuteron
breakup, was suppressed by cuts on the energy loss and time of
flight of charged particles in the Fd detector.
Figure.~\ref{fig4}(a) shows the energy loss times the square of
the particle velocity versus the laboratory momentum. A
projection of these events onto the ordinate gives a clear peak
with a low background. In our analysis we used a very loose cut
at the $6\sigma$ level but, as will be shown in
Sec.~\ref{sec:EtaMass}, changing this to $3\sigma$ would have
only a tiny effect on the value obtained for $m_{\eta}$.

Figure.~\ref{fig4}(b) shows a typical example for one
scintillation counter combination of the length of the
reconstructed flight path of the $^3$He selected from the
events in Fig.~\ref{fig4}(a). This has been calculated from the
TOF information between the first and the last scintillation
wall in the forward detector and the reconstructed particle
momentum, assuming the mass and charge of the $^3$He. A clear
peak is again evident with only a moderate background. Reducing
the $3\sigma$ cut used here to $2\sigma$ gives an even smaller
change in the value of $m_{\eta}$.

\begin{figure}[htb]
\centering
\includegraphics[width=1.0\linewidth]{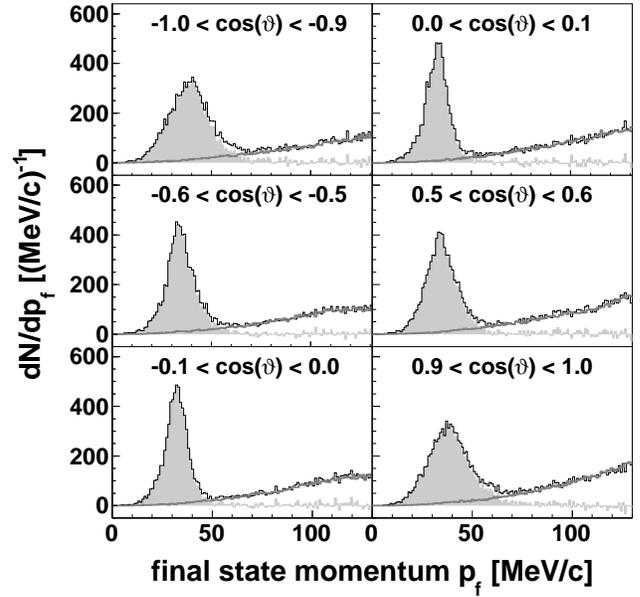}
\caption{\label{fig5} Center-of-mass distributions of the $^3$He momentum
from the $dp\rightarrow\,^{3}\text{He}\,X$ reaction for six
typical polar angle bins at the lowest excess energy measured,
$Q = 1.1$~MeV. The experimental data summed over $\phi$ are
shown by black lines and the background estimated from
subthreshold data by gray lines. The resulting
background-subtracted $dp\rightarrow\,^{3}\text{He}\,\eta$
signal is shaded gray.}
\end{figure}

The remaining background, shown in Figs.~\ref{fig3} and
\ref{fig5}, originates mainly from some residual deuteron
breakup and multipion production in the
$dp\rightarrow\,^{3}\text{He}\,X$ reaction, where
$X=(\pi\pi)^0$ or $X=(\pi\pi\pi)^0$. At the lowest excess
energy, the signal/background ratio is around 11 but this
decreases with increasing excess energy to $\approx 1.8$ at $Q
= 10.4$~MeV. This background was subtracted using the data
taken below the $\eta$ threshold at an excess energy of $Q
\approx -5$~MeV. These data were kinematically transformed to
positive $Q$ in order to compare them with results obtained
above threshold. The details of this technique are described
for missing-mass spectra in Ref.~\cite{MER2007}, but the method
is equally applicable to final-state momentum spectra. Due to
the very high statistics of the current experiment, the
distribution in $p_f$ could be investigated for twenty bins
each in $\vartheta$ and $\phi$. This is illustrated in
Fig.~\ref{fig5}, where examples of the $p_f$ spectra summed
over $\phi$ are shown for six $\cos\vartheta$ bins for the
energy closest to threshold, $Q = 1.1$~MeV. A similar picture
is found for the $\phi$ dependence after summing over $\theta$.

Mean values of the $^3$He momentum $p_f$ and the peak widths
for different $\vartheta$ and $\phi$ bins were extracted from
the background-subtracted $dp\rightarrow\,^{3}\text{He}\,\eta$
distributions by making Gaussian fits. A variation of the width
of 4--12~MeV/$c$ (rms) was found, as well as a displacement of
the mean value, both of which depended upon the polar and
azimuthal angles. This striking effect results from the
different resolutions of the ANKE Fd-system in $p_x$, $p_y$ and
$p_z$, which are discussed in the following section.

\subsection{Influence of the momentum resolution on the reconstructed final-state momentum}
\label{subsec:FM2}

\begin{figure}[hbt]
\flushright
\includegraphics[width=1.0\linewidth]{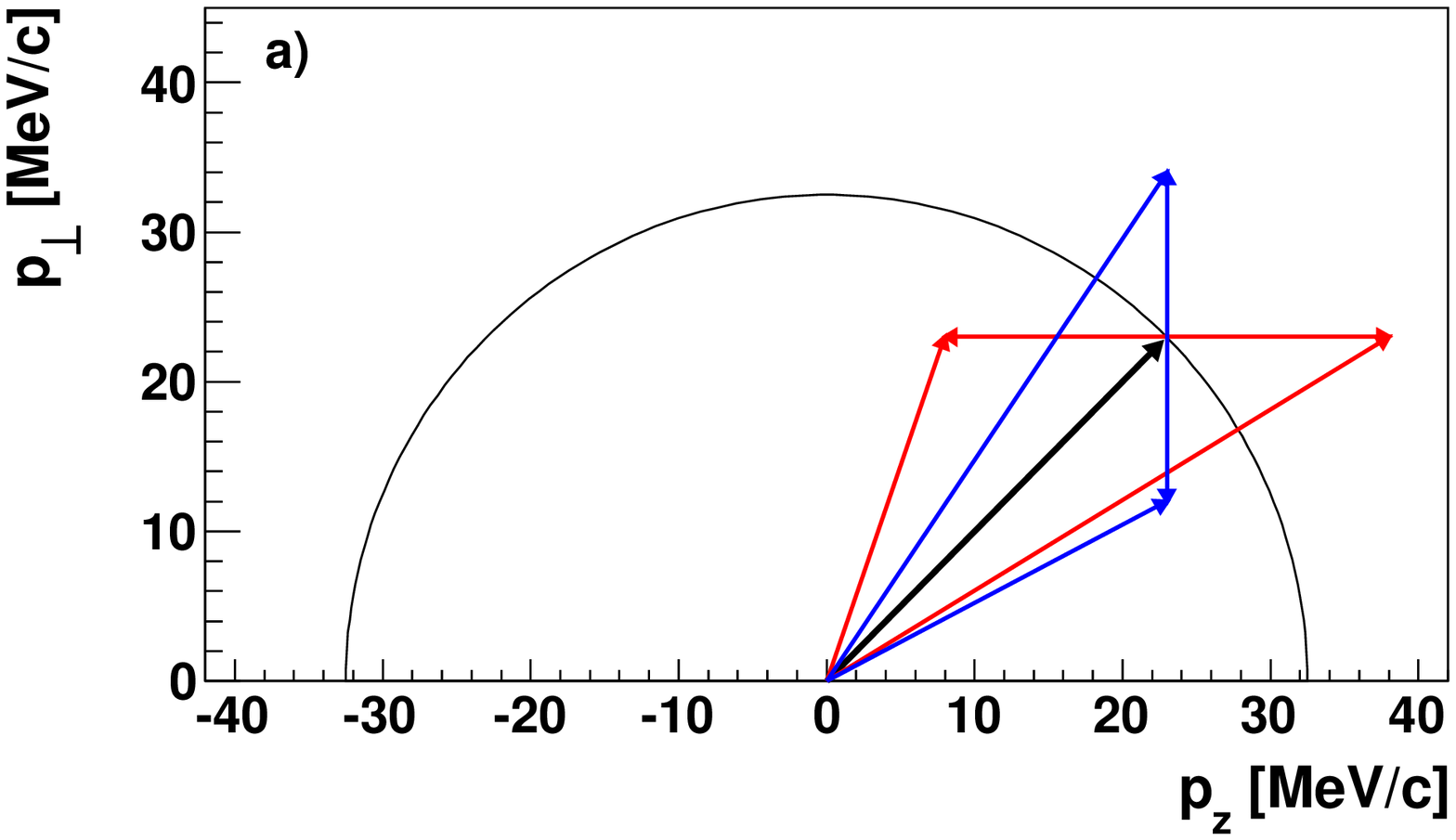} \\
\includegraphics[width=1.0\linewidth]{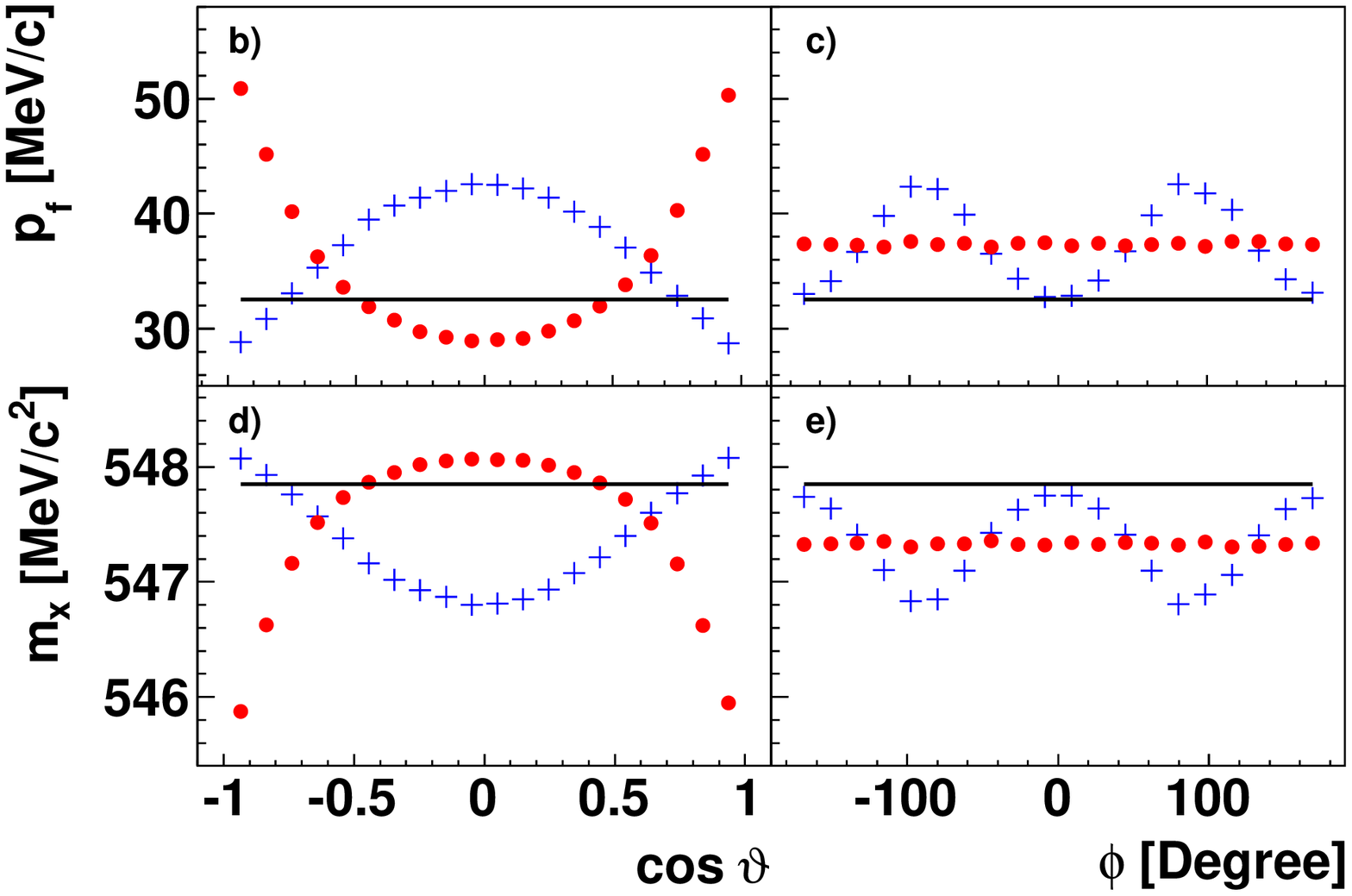} \\
\caption{\label{fig6} (color online) Influence of resolution on
the determination of the final-state momentum. The ideal $p_f$
sphere of panel (a) (black) is changed along the horizontal
(red) arrow by finite resolution in the longitudinal ($z$)
direction, whereas resolution effects in the transverse
direction are indicated by the vertical (blue) arrow. Panels
(b) -- (e) show the possible distortions at $Q=1.1$~MeV,
evaluated in Monte Carlo simulation. The mean values of both
the final-state momentum and the missing-mass distributions for
individual $\cos\vartheta$ and $\phi$ bins are shown without
(black line) and with momentum smearing in the $z$-direction
(red circles) and transversely (blue crosses). }
\end{figure}

The influence of resolution on the angular dependence of the
reconstructed $^3$He momentum and the missing-mass
distributions for the $dp\rightarrow\,^{3}\text{He}\,\eta$
reaction is illustrated by the two-dimensional $(p_{\bot},p_z)$
projection shown in Fig.~\ref{fig6}(a). In the ideal case of a
measurement with perfect resolution, the final-state momenta
are distributed on a sphere of constant radius $p_f$, as
indicated by the black line. Both the missing mass and
final-state momentum are then independent of $\cos\vartheta$
and $\phi$, as illustrated by the black line in
Figs.~\ref{fig6}(b)-6(e).

In a real experiment, the reconstructed momenta in the
laboratory frame are smeared by the finite resolution
associated with the detector setup and reconstruction
algorithms. If only $p_z$ were smeared, with a Gaussian width
of say $\sigma_z=30$~MeV/$c$ in the laboratory frame, an event
on the momentum sphere, indicated by the black arrow in
Fig.~\ref{fig6}(a), could be shifted along the red arrows. It is
important to note that for $p_f^{\text{rec}} <
p_f^{\text{true}}$ events are shifted toward lower
$|\cos\vartheta|$ whereas the reverse is true for
$p_f^{\text{rec}} > p_f^{\text{true}}$. This effect leads to a
$p_f$ distribution that is a function of $\cos\vartheta$
(Fig.~\ref{fig6}(b), red circles), i.e., the momentum sphere is
stretched for large longitudinal momenta and compressed for
high transverse momenta. For simple kinematic reasons, the
missing mass shows the inverse behavior [see Fig.~\ref{fig6}(d)].
Since the smearing was here assumed to be independent of $p_x$
and $p_y$, and hence $\phi$, this is reflected in the constancy
of the reconstructed momentum in Fig~\ref{fig6}(c). Nevertheless,
its value is higher than the true one, $p_f^{\text{true}}$.

When only the transverse momenta are smeared, but with
different Gaussian widths, e.g., $(\sigma_x, \sigma_y,
\sigma_z) = (10,20,0)$~MeV/$c$, the reconstructed momentum has
the opposite dependence on $\cos\vartheta$. As shown by the
blue crosses in Fig.~\ref{fig6}(b), $p_f$ decreases for
$\cos\vartheta \approx \pm 1$ and increases for $\cos\vartheta
\approx 0$. Different resolutions in $p_x$ and $p_y$ also
introduce a dependence on $\phi$, and this leads to
oscillations in both $p_f^{\text{rec}}$ and the missing mass in
the plots of Figs.~\ref{fig6}(c) and 6(e). The amplitude and phase of
these oscillations depend on the ratio $\sigma_x/\sigma_y$.

In reality, all three momentum components are reconstructed
with finite and generally different resolutions and the effects
described above will be superimposed, though they will be
dominated by the component with the worst resolution. The
determination of the $\eta$ mass has to take these kinematic
resolution effects into account because, without so doing, the
value extracted for $m_{\eta}$ would depend on the production
angle. For the current ANKE experiment, differences in
$m_{\eta}$ of up to 0.5~MeV/$c^2$ are found between
$\cos\vartheta = \pm 1$ and $\cos\vartheta = 0$. Furthermore,
the average of the final-state momentum over all
$\cos\vartheta$ and $\phi$ is shifted to a higher value than
the true one and the missing mass shifted to a lower value. The
angular distribution of the
$dp\rightarrow\,^{3}\text{He}\,\eta$ reaction could modify
slightly the effects of the resolution but, since this
variation is linear in $\cos\vartheta$ over all the $Q$-range
studied~\cite{MER2007}, even this is of little consequence for
the determination of $m_{\eta}$, given the symmetry in
$\cos\vartheta$ shown in Fig.~\ref{fig6}(b).

Precise determinations of the resolutions in $(p_x, p_y, p_z)$
are absolutely essential in order to correct the measured
kinematic variables. How this is done, by demanding an
isotropic $p_f$ momentum sphere, will be discussed next in some
detail in Sec.~\ref{subsec:FM3}.

\subsection{Correction to the final-state momenta}
\label{subsec:FM3}

\begin{figure}[hbt]
\centering
\includegraphics[width=1.0\linewidth]{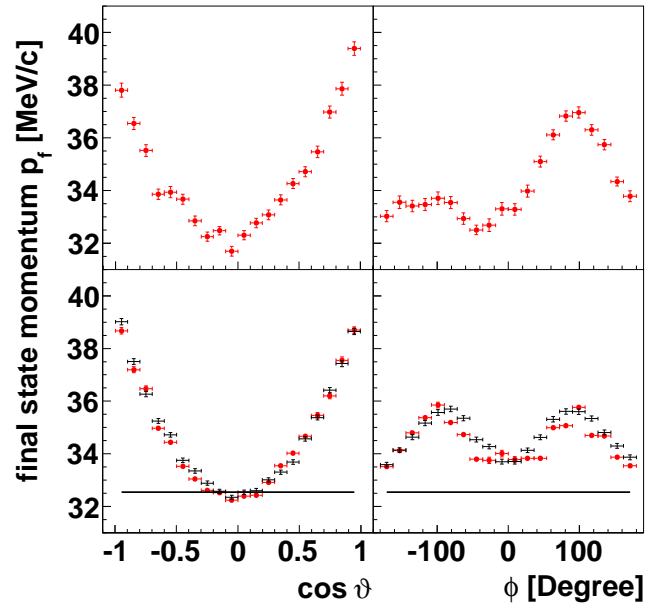}
\caption{\label{fig7} (color online) The mean values of the
final-state momentum distributions are shown for individual
$\cos\vartheta$ and $\phi$ bins for the standard (top) and
improved (bottom) calibration at $Q=1.1$~MeV (red circles). The
results of Monte Carlo simulations are shown without (black
horizontal line) and with momentum smearing (black points).
The comparison of the data with the simulation leads to a
determination of the momentum resolutions in the three
directions. }
\end{figure}

Mean values of the measured final-state momenta for
background-subtracted $dp\to{}^3\textrm{He}\,\eta$
distributions are shown in Fig.~\ref{fig7} for twenty
individual $\cos\vartheta$ and $\phi$ bins at an excess energy
$Q = 1.1$~MeV, before and after the improvement of the
calibration. The results for the standard calibration,
presented in the upper panels, show that the momentum sphere is
neither centered nor symmetric. The momentum sphere is shifted
to higher $p_z$, i.e., on average $p_f$ is higher for $^3$He
produced in the forward direction than in the backward. The
oscillations in the $\phi$ spectrum are also far from being
symmetric, and this is particularly evident at $\phi \approx
\pm 90^\circ$, where the $p_y$ momentum component dominates.
This asymmetric pattern is rather similar at all twelve
energies and this stresses the need to improve the calibration
for the determination of the correct momenta.

The $p_y$ component depends sensitively on the relative
$y$-position between the two wire chambers in the forward
detector and, by varying their positions by about 0.3~mm, the
momentum sphere could be centered in $p_y$. Changes in the
magnetic field strength of 0.0015~T at 1.4~T, i.e., changes of
the order of 0.1\%, allow one to make the necessary adjustment
in the $p_z$ component. The $p_x$ component can be fine-tuned
by adjusting the deflection angle $\alpha$ by about 0.4\% at
$5.8^{\circ}$. By making changes such that the mean values of
the final-state momenta are distributed on a centered and
perfectly symmetric sphere in $\cos\vartheta$ and $\phi$, the
detector alignment can be significantly improved, as shown in
the lower part of Fig.~\ref{fig7}. This procedure was carried
out using the data at all twelve energies simultaneously. The
magnitudes of these changes are so small that they have no
impact on the values of the missing masses of the different
reactions used in the standard calibration.

The improved spectra, shown in the lower half of
Fig.~\ref{fig7} for one of the twelve energies, allow one to
study the momentum smearing in the three directions. The values
of $\sigma_x$ and $\sigma_y$ were determined from the amplitude
and phase of the oscillation in $\phi$, as explained in
Sec.~\ref{subsec:FM2}, whereas that of $\sigma_z$ was extracted
by making a second order fit to the data and simulations for
$p_f$ as a function of $\cos\vartheta$. An additional
constraint is that the values of $(\sigma_x,\sigma_y,\sigma_z)$
must reproduce the width of the $^3$He$\, \eta$ final-state
momentum signal when integrated over all $\vartheta$ and
$\phi$. The resolution parameters were determined from the
spectra with uncertainties of $(\Delta \sigma_x,\Delta
\sigma_y,\Delta \sigma_z) = (0.2,\, 0.2,\, 0.1)$~MeV/$c$.

\begin{figure}
\centering
	\includegraphics[width=1.0\linewidth]{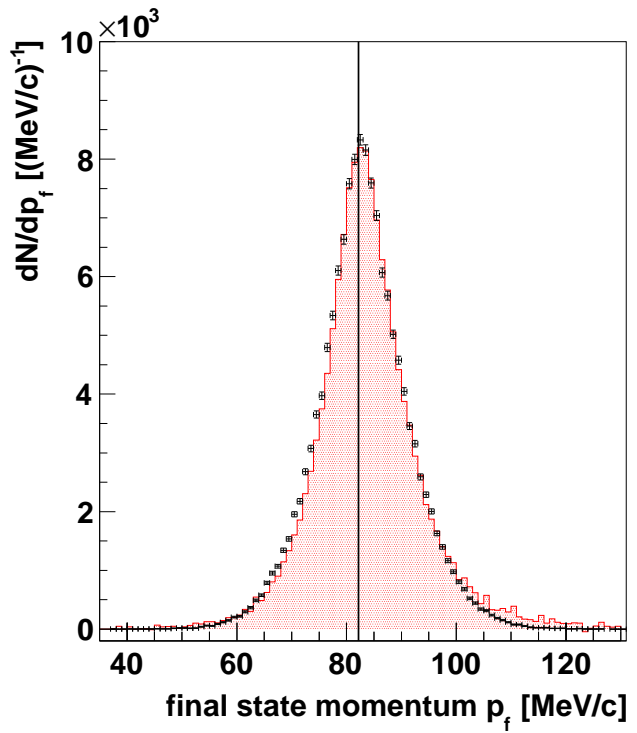}
\caption{\label{fig8} (color online) Final state momentum
distribution for background-subtracted $dp\to{}^{3}$He$\, \eta$
data (filled red area) and simulation (black crosses) for all
events at an excess energy of $Q\approx 8.6$~MeV. The vertical
line indicates the true final-state momentum assumed in the
simulation.}
\end{figure}

The individual momentum spreads were determined separately for
each of the twelve energies. In contrast to $\sigma_z$, which
is constant to within 1~MeV/$c$, $\sigma_x$ decreases and
$\sigma_y$ increases with excess energy. This behavior is
reasonable because the cone angle of the $^3$He ejectiles from
the $dp\to{}^{3}$He$\, \eta$ reaction increases with excess
energy. Close to threshold the hit positions in the chambers
are located in a small area near the center of the wire
chambers, whereas at higher $Q$ the hits are more widely
distributed. With the resolution parameters thus determined,
data similar to those shown in Fig.~\ref{fig7} are described
well by Monte Carlo simulations at all twelve energies.
Figure.~\ref{fig8} shows the distribution of the final-state
momentum summed over all $\cos\vartheta$ and $\phi$ at an
excess energy of $Q\approx 8.6$ MeV. It is clear that the
measured momentum distribution (filled area) is generally very
well reproduced by the Monte Carlo simulated data (black
crosses).

Using data from all twelve energies above threshold, a mean
momentum resolution of ANKE was calculated in the narrow $^3$He
laboratory momentum range of 2.63--2.68~GeV/$c$. The values of
the momentum spreads in the laboratory frame were found to be
$(\sigma_x,\sigma_y,\sigma_z) = (2.8,7.9,16.4)$~MeV/$c$. As
expected for a fixed target experiment, the resolution in $p_z$
is by far the poorest. Furthermore, because of the particular
construction of the wire chambers, the $p_x$ resolution is
better than that for $p_y$. However, in the determination of
the final-state momentum corrections, the individual resolution
parameters determined at each beam momentum were used.

\begin{figure}
\centering
	\includegraphics[width=1.0\linewidth]{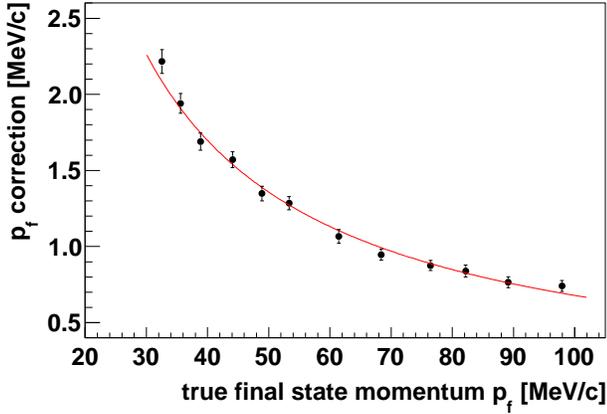}
\caption{\label{fig9} (color online) Deviation of the measured
final-state momentum in the $dp \to {}^3\textrm{He}\,\eta$
reaction from the true one due to resolution effects, evaluated
in Monte Carlo simulation. The twelve measured final-state
momenta in the near-threshold region, $Q =$~1--11~MeV, have to
be corrected by 0.7--2.2~MeV/$c$ to compensate for such
effects. The curve is a $1/p_f$ fit to the points.}
\end{figure}

Owing to the resolution effects shown in the lower half of
Fig.~\ref{fig7}, the average of the final-state momentum over
all $\cos\vartheta$ and $\phi$ is shifted to a higher value
than the true one (black horizontal line). By comparing the
averages resulting from the Monte Carlo simulations with and
without momentum smearing, correction parameters were
calculated for all twelve energies and these are presented in
Fig.~\ref{fig9}. The correction is about 2.22~MeV/$c$ for the
lowest momentum and decreases steadily with $p_f$. The error
bars shown are dominated by the uncertainties in the resolution
parameters $(\Delta \sigma_x,\Delta \sigma_y,\Delta \sigma_z)$
and range from 0.08~MeV/$c$ at the lowest beam momentum to
0.04~MeV/$c$ at higher energies. It should be noted that the
dependence of the correction parameters on the value assumed
for the $\eta$ mass is negligible.

If the resolution factors $\sigma_i$ are largely independent of
the beam momentum, the correction should vary like $\sim
1/p_f$. This behavior arises because the deviation depends on
the ratio of the ANKE momentum resolution to the size of the
momentum sphere. Confirmation of such a dependence is offered
by the curve, which is a $1/p_f$ fit to the data. Despite the
good $\chi^2/\text{NDF} \approx 0.9$, the individual Monte
Carlo estimates were used in the subsequent analysis.

Compensation for the effects of the momentum resolution is
essential for an accurate determination of the production
threshold. Without this correction, the value obtained for the
$\eta$ mass would be lower by about 150~keV/$c^2$.

\section{The mass of the $\boldsymbol{\eta}$ meson}
\label{sec:EtaMass}

In order to obtain a robust value for the mass of the $\eta$
meson, it is necessary to extrapolate the experimental data
with high precision in order to determine the value of the
deuteron beam momentum at threshold. For this purpose we show
the complete set of final-state and beam momenta $(p_f,p_d)$
with statistical uncertainties in Table~\ref{tab1:pfpd}. Also
shown are the corrections that were included in the
values quoted for $p_f$.

Due to the resolution effects, the twelve $p_f$ distributions
are found to be slightly asymmetric over the whole angular
range, both for the experimental data as well as the Monte
Carlo simulated events, as shown for a typical energy in
Fig.~\ref{fig8}. The values of the $p_f$ means were determined
for both data and simulations within $\pm 3\sigma$ limits. The
good statistics of $\approx 1.3 \times 10^{5} \
^{3}\rm{He}\,\eta$ events for each energy meant that the
uncorrected value for $p_f$ could be extracted with an
uncertainty of $\approx 23$~keV/$c$. In total $\approx 1.5
\times 10^{6} \ ^{3}\rm{He}\,\eta$ events were collected in the
experiment. After applying the resolution correction to the
measured $p_f$, the overall uncertainty is increased, as
indicated in Table~\ref{tab1:pfpd}.

Although the description of the reconstructed final-state
momentum distribution in Fig.~\ref{fig8} by the Monte Carlo
simulation is very good, it is not perfect, especially in the
high momentum tail. Such discrepancies could arise from slight
imperfections in the spectrometer calibration, $^3$He
scattering in the wire chambers, or limitations in the
background subtraction approach. In order to quantify their
influence on the value extracted for the $\eta$ mass, the
interval used to determine the means of the $p_f$ was varied
between $\pm 2\sigma$ to $\pm 4\sigma$, where $\sigma$
represents the peak width. Such a variation leads to a
collective shift in the extracted final-state momenta of
approximately 0.16~MeV/$c$. Since this effect corresponds to
an overall shift in the final state momenta, it is not included
in the numbers quoted in Table~\ref{tab1:pfpd} but must be
considered in the final $\eta$ mass determination, where it
introduces a systematic uncertainty of 12~keV/$c^2$.

\begin{table}[hbt]
\caption{\label{tab1:pfpd} Values of the laboratory beam
momenta $p_d$, corrected final-state c.m. momenta $p_f$, and the
$p_f$ correction parameters measured at twelve different excess
energies; the statistical uncertainties are noted in brackets.
The approximate values of $Q$ quoted here are merely used to
label the twelve settings.} \vspace{2mm}
\begin{tabular}{c c c c}
\hline
Excess 			& Beam 								& Final-state						& $p_f$ correction\\			
energy $Q$	& momentum $p_d$				& momentum $p_f$ 			& parameter\\
MeV				&	MeV/$c$							& 	MeV/$c$ 						& MeV/$c$\\
\hline
1.1				&		$3146.41(3)$		&		$32.46(8)$	&	$2.22(8)$\\
1.4				&		$3147.35(3)$		&		$35.56(7)$	&	$1.94(6)$\\
1.6				&		$3148.45(3)$		&		$39.00(6)$	&	$1.69(6)$\\
2.1				&		$3150.42(3)$		&		$44.09(6)$	&	$1.57(5)$\\
2.6				&		$3152.45(3)$		&		$49.25(5)$	&	$1.35(5)$\\
3.1				&		$3154.49(3)$		&		$53.66(5)$	&	$1.28(4)$\\
4.1				&		$3158.71(3)$		&		$61.70(5)$	&	$1.07(5)$\\
5.1				&		$3162.78(3)$		&		$68.77(4)$	&	$0.95(4)$\\
6.3				&		$3168.05(3)$		&		$76.92(4)$	&	$0.88(3)$\\
7.3				&		$3172.15(3)$		&		$82.64(5)$	&	$0.84(4)$\\
8.6				&		$3177.51(3)$		&		$89.81(4)$	&	$0.76(4)$\\
10.4				&		$3184.87(3)$		&		$98.64(4)$	&	$0.74(4)$\\
\hline
\end{tabular}
\end{table}

\begin{figure}[hbt]
\includegraphics[width=1.0\linewidth]{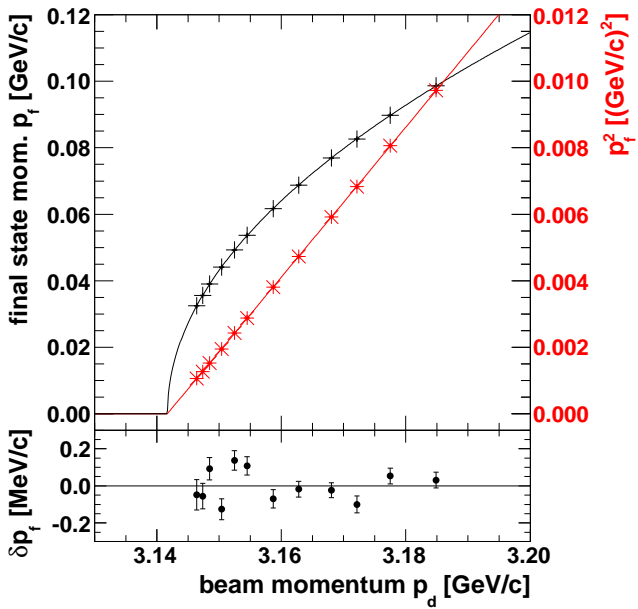}
\caption{\label{fig10} (color online) Corrected values of the
final-state c.m. momentum $p_f$ (black crosses) and its square
(red stars) plotted against the deuteron laboratory momentum
$p_d$. The error bars are too small to be shown on the figure.
The extrapolation to threshold is carried out on the basis of
Eq.~\eqref{eq:pf}, where a scaling factor $S$ has been
introduced. The lower panel shows the deviations of the
experimental data from the fitted curve in $p_f$. The errors
shown here do not include the overall systematic uncertainty in
$p_f$ associated with the description of the profile in
Fig.~\ref{fig8}.}
\end{figure}

The extrapolation of the data to threshold is illustrated in
Fig.~\ref{fig10} for both $p_f$ and $p_f^{\,2}$ versus $p_d$.
Whereas, to first order, $p_f^{\,2}$ depends linearly on $p_d$,
the analysis considers the full dependence $p_f = p_f(m_{\eta},
S, p_{d})$, as given by Eqs.~\eqref{eq:pf} and \eqref{eq:s}.
Only the $\eta$ mass, chosen as a free parameter, defines the
production threshold. The scaling factor $S$, discussed in
Sec.~\ref{sec:Method}, allows for a possible systematic energy
dependence of $p_f$. This would represent yet a further fine
tuning of the description of the measurement process but it is
crucial to note that its introduction does not affect the value
obtained for the threshold momentum and hence $m_{\eta}$.

The overall fit to the data in Fig.~\ref{fig10} has a
$\chi^2/\text{NDF}=1.28$ and the best value of the mass quoted
in Table~\ref{tab2:EtaMass} is
$m_{\eta}=(547.873\pm0.005)$~MeV/$c^2$, where the error is
primarily statistical. The corresponding deuteron momentum at
threshold is $p_d = (3141.688\pm0.021)$~MeV/$c$. A linear fit of
$p_f^{\,2}$ versus $p_d$ would give a poorer reduced $\chi^2$
and a mass that was 10~keV/$c^2$ higher.

\begin{table}[hbt]
\caption{\label{tab2:EtaMass} Values of the $\eta$-mass and
scaling factor evaluated separately for the two supercycles and
for the complete data set. The errors do not include the
systematic uncertainties in the determination of the beam
momentum.} \vspace{2mm}
\begin{tabular}{c c c}
\hline
Supercycle	&	Scaling factor $S$		&	$m_{\eta}$ \\
\hline
1			&	$1.008 \pm 0.001$	&	$(547.870\pm0.007)$~MeV/$c^2$ \\
2			&	$1.008 \pm 0.001$	&	$(547.877\pm0.007)$~MeV/$c^2$ \\
1+2  	    &	$1.008 \pm 0.001$	&	$(547.873\pm0.005)$~MeV/$c^2$ \\
\hline
\end{tabular}
\end{table}

The scaling factor $S=1.008\pm 0.001$ is well determined and
differs very slightly from unity, which means that the twelve
momentum spheres are about 0.8\% bigger than expected. One
consequence of this is that the missing mass is not constant
and one would get a slightly different value at each of
the twelve energies studied. If one took $S=1.0$, the
$\chi^2/\text{NDF}$ would jump to 24.7 and this would result
in a shift of 64 keV/$c^2$ in the $\eta$ mass. Such a
systematic error is avoided in the threshold-determination
method described in Sec.~\ref{sec:Method}.

By far the dominant systematic errors arise from the
determinations of the absolute value of the beam momentum and
the $p_f$ correction parameters. As can be seen from
Table~\ref{tab3:Systematics}, all other sources, such as
effects from the time stability of the data, further
contributions from the fine calibration, the event selection,
the background subtraction for the $p_f$ distributions, as well
as contributions of the $\eta$ mass assumed in Monte Carlo
simulations, are negligible in comparison.

\begin{table}[hbt]
\caption{\label{tab3:Systematics} Systematic uncertainties in
the determination of $m_{\eta}$. The small ``experimental
settings'' contribution includes effects from the magnetic
field, the deflection angle, and the (vertical) wire chamber
positions, all of which are coupled. The PDG value of
$m_{\eta}$~\cite{PDG2010} was used in the simulations but, if
our result were used, it would only result in a 2~keV/$c^2$
change. The effects of putting stricter cuts on $\Delta E
\times \beta^2$ and the flight length are also shown.}
\vspace{2mm}
\begin{tabular}{l l r}
\hline
Source											&		Variation\hspace{2mm}			& $\Delta m_{\eta}$ \\
&	&  keV/$c^2$  	\\
\hline
Absolute beam momentum\hspace{4mm} 	&		95 keV/$c$									&	23 			\\
Experimental settings									&															&	2 	 			\\
$m_{\eta}$ assumed in simulations				&		20 keV/$c^2$								&	$< 2$		\\
$\Delta E \times \beta^2$ cut						&		$6\sigma \rightarrow 2\sigma$		&	5				\\
Flight length cut											&		$3\sigma \rightarrow 2\sigma$		&	1				\\
$p_f$ correction parameters						&		$4\sigma \rightarrow 2\sigma$		& 	12				\\
\hline
Total systematic uncertainty							&															& 27		\\
\hline
\end{tabular}
\end{table}

This uncertainty in the beam momentum translates into one in
the mass of
\begin{equation}
\label{eq:uncertainty}
\Delta m_{\eta} = \frac{m_pp_d}{(m_{^3\text{He}}+m_{\eta})E_d}\,\Delta p_d
=23~\text{keV}/c^2,
\end{equation}
and hence, taken together with all other systematic
uncertainties, to a final value of
\begin{equation}
m_{\eta}=(547.873\pm0.005_{\text{stat}}\pm0.027_{\text{syst}})~\text{MeV}/c^2.
\end{equation}

To investigate further some of the systematic effects, the
results were extrapolated separately for the data obtained in
the two supercycles and the individual values of the $\eta$
mass and the scaling factor $S$ are given in
Table~\ref{tab2:EtaMass}. There is only a tiny difference
between the two separately determined $\eta$ mass values of
7~keV/$c^2$. This agreement supports the validity of taking the
mean values of the beam momenta determined at the beginning and
end of the measurements and the subsequent correct handling of
the corresponding systematic uncertainties.

\section{Conclusions}
\label{sec:Conc}

\begin{figure}[hbt]
\includegraphics[width=1.0\linewidth]{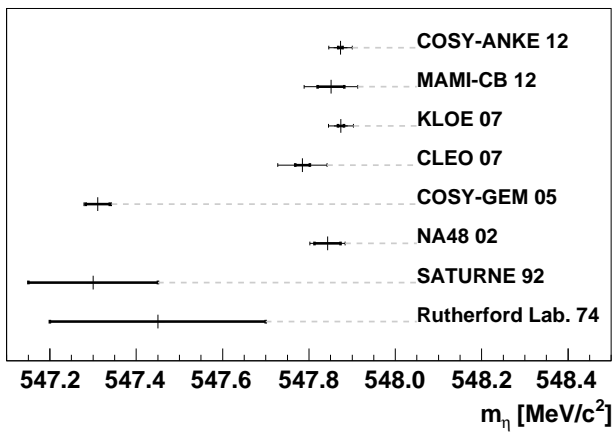}
\caption{\label{fig11} Results of the different $\eta$ mass
experiments. Where two error bars are shown, the heavy line
indicates the statistical uncertainty and the faint ones the
systematic. The earlier missing-mass experiments, marked
Rutherford Lab.\ 74~\cite{DUA1974}, SATURNE 92~\cite{PLO1992},
and COSY-GEM 05~\cite{ABD2005}, all obtained low values of
$m_{\eta}$ compared to the experiments where the meson was
identified through its decay products, viz.\ NA48
02~\cite{LAI2002}, KLOE 07~\cite{AMB2007}, CLEO
07~\cite{MIL2007}, and MAMI-CB 12~\cite{MAMI2012}. Our result,
COSY-ANKE 12, is completely consistent with these more refined
experiments.}
\end{figure}

We have measured the mass of the $\eta$ meson in a missing-mass
experiment by identifying precisely the production threshold in
the $dp\to{}^{3}$He$\,\eta$ reaction. As is seen in
Fig.~\ref{fig11}, the value obtained is consistent with all the
recent measurements where the meson decay products were
studied~\cite{LAI2002,AMB2007,MAMI2012,MIL2007}. The precision
achieved is similar to these works and the deviation from the
PDG best value~\cite{PDG2010} is only 20~keV/$c^2$, which is
less than our systematic error.

This success was based upon a precise determination of the beam
momentum using the spin-resonance technique~\cite{GOS2010}, a
clear identification of the $\eta$ signal~\cite{MER2007}, and a
systematic study of the measurement of the $^3$He$\, \eta$
final-state momentum in the ANKE spectrometer. The latter was
made possible only through the complete geometric acceptance of
ANKE for the $dp \to {}^3\textrm{He}\,\eta$ reaction close to
threshold. This allowed us to require that the c.m. momentum in
the final state should be identical in all directions. This is
a powerful technique that might be useful for other two-body
reactions. Unlike the MAMI methodology~\cite{MAMI2012}, the
experiment relied purely upon kinematics rather than yields to
determine the threshold momentum and hence the meson mass.
However, it is important to realize that the anomalous behavior
of the production cross section, where the cross section jumps
so rapidly with excess energy, leads to the desirable high
count rates near threshold.

Our result differs by about 0.5~MeV/$c^2$ from earlier
missing-mass evaluations~\cite{DUA1974,PLO1992,ABD2005} and so
the hypothesis of a distortion of the background under the
$\eta$ peak must be discarded. Unlike experiments with external
targets, the energy loss in a windowless cluster jet is
negligible, but the current experiment had other advantages
over these measurements. In particular, if we had access only
to data taken in the forward direction, this would not have
allowed the fine tuning of the spectrometer and a somewhat
different value would have been found for $m_{\eta}$.

Finally, twelve energies above threshold were investigated and
this allowed a reliable extrapolation to be made to find the
production threshold. This is intrinsically subject to far
fewer systematic uncertainties than a measurement at a single
energy. It is therefore clear that, with care, a missing-mass
approach can be competitive with experiments in which meson
decays are measured.\\

\begin{acknowledgments}
The authors wish to express their thanks to the other members
of the COSY machine crew for producing such good experimental
conditions and also to the other members of the ANKE
collaboration for diverse help in the experiment. This work was
supported in part by the JCHP (Fremde Forschungs und
Entwicklungsarbeiten) FFE.
\end{acknowledgments}



\begin{thebibliography}{99}
%
\bibitem{LAI2002} A.~Lai \emph{et al.}, Phys.\ Lett.\ B
    \textbf{533}, 196 (2002).
%
\bibitem{AMB2007} F.~Ambrosino \emph{et al.}, J.\ High Energy
    Phys.\ \textbf{12}, 73 (2007).
%
\bibitem{MAMI2012} A.~Nikolaev, Ph.D thesis, University of Bonn
    (2012), available
    from \url{http://hss.ulb.uni-bonn.de/2012/2764/2764.htm};
    MAMI-CB experiment, submitted for publication.
%
\bibitem{MIL2007} D.~H.~Miller \emph{et al.}, Phys.\ Rev.\
    Lett.\ \textbf{99}, 122002 (2007).
%
\bibitem{PDG2010} K.~Nakamura \emph{et al.}\ (Particle Data
    Group), J.\ Phys.\ G \textbf{37}, 075021 (2010).
%
\bibitem{DUA1974} A.~Duane \emph{et al.}, Phys.\ Rev.\ Lett.\
    \textbf{32}, 425 (1974).
%
\bibitem{PLO1992} F.~Plouin \emph{et al.}, Phys.\ Lett.\ B
    \textbf{276}, 526 (1992).
%
\bibitem{ABD2005} M.~Abdel-Bary \emph{et al.}, Phys. Lett. B
    \textbf{619}, 281 (2005).
%
\bibitem{MER2007} T.~Mersmann \emph{et al.}, Phys.\ Rev.\
    Lett.\ \textbf{98}, 242301 (2007).
%
\bibitem{RAU2009} T.~Rausmann \emph{et al.}, Phys.\ Rev.\
    C \textbf{80}, 017001 (2009).
%
\bibitem{GOS2010} P.~Goslawski \emph{et al.}, Phys.\ Rev.\ ST
    Accel.\ Beams \textbf{13}, 022803 (2010).
%
\bibitem{NIST} NIST - National\ Institute\ for\ Standards\ and\ Technology,
\url{http://physics.nist.gov/cuu/Constants/index.html}.
%
\bibitem{DER1980} Ya.~S.~Derbenev \emph{et al.},
    Part.\ Accel.\ \textbf{10}, 177 (1980).
%
\bibitem{MAI1997} R.~Maier \emph{et al.}, Nucl.\ Instrum.\
    Methods Phys.\ Res.\ Sect. A \textbf{390}, 1 (1997).
%
\bibitem{BAR2001} S.~Barsov \emph{et al.}, Nucl.\ Instrum.\
    Methods Phys.\ Res.\ Sect. A \textbf{462}, 364 (2001).
%
\bibitem{KHO1999} A.~Khoukaz \emph{et al.}, Eur.\ Phys.\
    J.\ D \textbf{5}, 275 (1999).
%
\bibitem{CHI2002} B.~Chiladze \emph{et al.}, Phys.\ Part.\ Nucl.\
\textbf{113}, 95 (2002).
%
\bibitem{DYM2004} S.~Dymov \emph{et al.}, Phys.\ Part.\ Nucl.\
\textbf{119}, 40 (2004).
%
\bibitem{DYM2009} S.~Dymov, ANKE Notes No. 22: FD Momentum Calibration for March'08 Beam Time (2009),
    available from \url{www2.fz-juelich.de/ikp/anke/en/internal.shtml}.
%
\bibitem{KHO2007} A.~Khoukaz, COSY Proposal + Beam Request No. 187, available from 
    \url{www2.fz-juelich.de/ikp/anke/en/proposals.shtml}.

 \end{thebibliography}
\end{document}